\magnification=\magstep1
\hsize=16truecm
\vsize=22.2truecm
\baselineskip=12pt
\hfuzz=18pt
\parindent=1truecm
\parskip=0.27truecm

\font\srm=cmr7 scaled\magstep1

\font\sit=cmti7 scaled\magstep1
\def\sqr#1#2{{\vcenter{\vbox{\hrule height.#2pt
        \hbox{\vrule width.#2pt height#1pt \kern#1pt
           \vrule width.#2pt}
        \hrule height.#2pt}}}}

%
\def\lsim{{\displaystyle
{{\raise-8pt\hbox{$ <$}}
\atop{\raise5pt\hbox{$\sim$}}}}}
\def\gsim{{\displaystyle
{{\raise-8pt\hbox{$ >$}}
\atop{\raise5pt\hbox{$\sim$}}}}}
%
\def\slsim{{\displaystyle
{{\raise-8pt\hbox{$\scriptstyle <$}}
\atop{\raise5pt\hbox{$\scriptstyle \sim$}}}}}
\def\sgsim{{\displaystyle
{{\raise-8pt\hbox{$\scriptstyle  >$}}
\atop{\raise5pt\hbox{$\scriptstyle \sim$}}}}}
\def\Tr{\,{\rm Tr}\, }
\def\Im{\,{\rm Im}\, }
\def\Re{\,{\rm Re}\, }
\def\np#1#2#3{Nucl. Phys. {\bf{B#1}} (#2) #3}
\def\pl#1#2#3{Phys. Lett. {\bf{#1B}} (#2) #3}
\def\prl#1#2#3{Phys. Rev. Lett. {\bf{#1}} (#2) #3}

\def\nl{\hfil\break}
%
\newskip\oneline \oneline=1em plus.3em minus.3em
\newskip\halfline \halfline=.5em plus .15em minus.15em
\newbox\sect
\newcount\eq
\newbox\lett
\newdimen\short
\def\adv{\global\advance\eq by1}
\def\set#1#2{\setbox#1=\hbox{#2}}
\def\nextlet#1{\global\advance\eq
by-1\setbox\lett=\hbox{\rlap#1\phantom{a}}}
\newcount\eqncount\newcount\sectcount\eqncount=0\sectcount=0
\def\sectadv{\global\advance\sectcount by1}
\def\secta{\global\advance\sectcount by1}
\def\equn{\global\advance\eqncount
by1\eqno{(\copy\sect\the\eqncount)}}
\def\equnal{\global\advance\eqncount
by1{(\copy\sect\the\eqncount)}}
\def\put#1{\global\edef#1{\the\eqncount}}
\def\un{1}
\def\deux{2}
\def\trois{3}
\def\quatre{4}
\def\sept{5}
\def\huit{6}
\def\cinq{7}
\def\unif{8}
\def\treize{9}
\def\neuf{10}
\def\dix{11}
\def\six{12}
\def\onze{13}
\def\douze{14}
\def\thgen{15}
\def\quinze{16}
\def\seize{17}
\def\dixsept{18}
\def\quatorze{19}
\def\sch{20}
\def\vingt{21}
{\nopagenumbers
\line{\hfil CERN-TH/96-09 }
\line{\hfil hep-th/9605012}
\vskip 2cm
\centerline{\bf ONE-LOOP CORRECTIONS TO COUPLING CONSTANTS}
\centerline{\bf IN STRING EFFECTIVE FIELD
THEORY\footnote{$^{\,\diamond}$}
{\srm To appear in the proceedings of the {\sit 5th Hellenic
School and Workshops on Elementary Particle Physics,}
Corfu, Greece, 3--24 September 1995.}}
\vskip 2.8cm
\centerline{ P.M. PETROPOULOS\footnote{$^{\,\star}$}{\srm On leave
from {\sit Centre National de
la Recherche Scientifique,} France.}}
\vskip 0.3cm
\centerline{\it Theory Division, CERN}
\centerline{\it 1211 Geneva 23, Switzerland}
\vskip 2.8cm
\centerline{\bf Abstract}
\vskip 0.3cm
In the framework of a recently proposed method for computing
exactly string amplitudes regularized in the
infra-red, I determine the one-loop correlators for auxiliary
fields in the symmetric $Z_2\times Z_2$ orbifold model.
The $D$-field correlation function turns out to give
the one-loop corrections for the gauge couplings,
which amounts to a string-theory supersymmetry Ward identity.
The two-point function for uncharged $F$ fields leads to the
one-loop renormalization of the moduli K\"ahler metric, and
eventually to the corrections for the Yukawa couplings.
\vfil
\noindent\line{CERN-TH/96-09\hfil}
\noindent\line{April 1996\hfil}
\eject}
\pageno=2

{\bf 1. Introduction}

String theories have been advocated as the best candidates for
unifying all known and yet to be discovered interactions. This
unification takes place at the Planck scale, which is the natural
scale for a theory that contains quantum gravity. However, in order
to make contact with low-energy physics and eventually to reach
phenomenological predictions, one needs a consistent description of
the massless degrees of freedom of the string and it is remarkable
that this part of the spectrum can be described in terms of an
effective field theory which, for superstring compactifications,
turns out to be a supersymmetric Yang--Mills theory coupled to
gravity.

The most general $N = 1$ supergravity action including up to two
derivatives is fully specified by its K\"ahler potential $K (z, \bar
z)$, its superpotential $W (z)$ and its gauge function $f (z)$ [\un].
These are respectively real and holomorphic functions of the
chiral-superfield scalar components. The K\"ahler potential
determines the matter kinetic terms while the superpotential and the
gauge function are related to the Yukawa and gauge couplings
respectively. One of the main tasks for the determination of the
low-energy string effective theory is the computation of these
functions in the fundamental theory. This is achieved by computing
adequate string amplitudes as functions of the vev's of gauge singlet
fields corresponding to the moduli of the internal manifold. In most
cases (including symmetric orbifolds) such a computation can be
performed at the level of the sphere and leads in particular to the
so-called string unification: both gauge and Yukawa couplings are
homogeneously related to the dilaton vev [\deux], which plays the
role of the string coupling constant. Of major importance are,
however, the string loop corrections, since they contain all the
information that one needs at low energies about the decoupled
infinite tower of massive string modes [\trois]. The effects of these
massive states are summarized in the threshold corrections, which are
non-trivial functions of the moduli and spoil the tree-level
relations among couplings [\quatre --\cinq]. The threshold
corrections are therefore important for the issue of string
unification below the Planck mass [\unif , \treize]. They may also
have implications in the problem of supersymmetry breaking.

There is no general method for computing string loop amplitudes. When
the space-time fields involved in these amplitudes are associated
with truly marginal world-sheet vertex operators, the approach
becomes more systematic. Classical background sources are introduced
for these fields that generate exactly marginal deformations. Hence,
if one can solve the vacuum amplitude for the deformed model, any
correlation function of these operators will be reached exactly,
including the back-reaction of all gravitationally coupled
fields, by taking derivatives with respect to the above sources. This
is possible, in general, in lattice compactifications whenever the
deformation corresponds to a Lorentz boost. However, even in those
cases, ambiguities related to the infra-red divergences may arise,
due to the on-shell formulation of string theory in the presence of
massless particles. In a recent article [\cinq] an interesting method
for computing unambiguously the string loop corrections has been
proposed. The procedure that is used consists of replacing flat
four-dimensional space-time with a suitably chosen curved one in a
way that preserves gauge symmetry, supersymmetry and modular
invariance, and with curvature that induces an infra-red cut-off.

In these notes I would like to illustrate the aforementioned methods
and compute the one-loop correction to the K\"ahler metric for some
moduli fields in a particular heterotic-string compactification: the
$Z_2 \times Z_2$
orbifold model. The one-loop correction to the K\"ahler metric, i.e.
the scalar wave-function renormalization, and its moduli dependance
are of prime importance in the determination of the Yukawa couplings.
In fact, as long as supersymmetry is unbroken, the superpotential
does not receive any loop correction [\neuf], and consequently the
only corrections to the Yukawa couplings are induced by the
wave-function renormalizations. Of course, there is experimental
evidence in favor of non-supersymmetric low-energy physics.
Nevertheless, assuming that the breaking of supersymmetry occurs at a
scale of the order of 1 TeV, non-supersymmetric corrections to
dimensionless parameters are not important, and there is some
relevance in computing corrections in the context of unbroken
supersymmetry.

In order to apply the background-field method successfully, I will
compute string vacuum amplitudes with insertions of $F$ auxiliary
fields associated with the untwisted moduli. The corresponding vertex
operators turn out to be truly marginal world-sheet operators. Their
classical background fields generate Lorentz boosts, and any of their
correlation functions is therefore exactly calculable. Thanks to the
presence of supersymmetry, one expects the correlation functions
determined in this way to be identical to those that would have been
obtained by using insertions of the moduli-field kinetic terms, and
therefore to lead to the moduli wave-function renormalization. Of
course, this statement is based on the existence of Ward identities,
which relate the renormalization of various members of a
supermultiplet and that
have not been strictly proved in string theory. However, in order to
show the consistency of my calculation, I will proceed to the
determination of amplitudes with insertions of $D$ auxiliary fields.
The result for the two-point function turns out to be in this case
identical to the one obtained in [\cinq] for the double insertion of
the magnetic field operator, and thus demonstrates a supersymmetry
Ward identity for the gauge multiplet.

These notes are organized as follows. Section 2 will be a reminder of
the
$Z_2 \times Z_2$
orbifold compactification of the heterotic string. I will essentially
settle the notations and explain how to compute exactly one-loop
string amplitudes of vertex operators associated with a marginal
background deformation corresponding to Lorentz boosts. In section 3,
I will proceed to the explicit determination of the two-point
function for the $D$ fields associated with the vector
supermultiplet. This will lead to the one-loop gauge coupling
correction. Section 4 will be devoted to an analogous computation for
the $F$ components of the untwisted moduli chiral superfields. I will
conclude on the possible generalizations of this work in the last
section.
\vskip 0.4cm
{\bf 2. Marginal background deformations and the $Z_2 \times Z_2$
orbifold}

The background field method has been extensively used in field theory
for the computation of various quantum corrections. It is easily
adapted to string theory where the aim is to determine, order by
order in the topological expansion, the low-energy effective theory
parameters. These parameters are associated with four-dimensional
fields that correspond to massless string excitations. Their
computation therefore amounts to the evaluation of string amplitudes
$$
\int_{{\cal F}_g} d\Omega_g \left\langle
{1\over 16\pi ^3}\int  \! 2i\, dz\,d{\bar z} \,
V_1(z,{\bar z})\;
{1\over 16\pi ^3}\int  \! 2i\, dz\,d{\bar z} \,
V_2(z,{\bar z})\, \ldots
\right\rangle_g                             \equn\put\eun
$$
at zero external momenta. Hence,
$V_j(z,{\bar z})$ are the internal conformal field theory part of
some massless vertex operators. In expression (\eun)
$d\Omega_g$ is the invariant measure for genus-$g$ moduli,
${\cal F}_g$ the corresponding fundamental domain and
$\langle\;\rangle_g$ denotes a conformal field theory correlation
function on a genus-$g$ surface. Vertices $V_j(z,{\bar z})$ are
weight-$(1, 1)$ operators, and correlators such as those appearing in
(\eun) are in principle calculable by switching on the corresponding
constant backgrounds $f_j$, namely by adding to the two-dimensional
action the deformation
$$
\Delta S=-{1 \over 16\pi ^3}\int \!2i\, dz\,d{\bar z} \,
\sum_j f_j \, V_j(z,{\bar z})\; ,           \equn\put\edeux
$$
and computing the deformed vacuum amplitude
$
Z\left({\bf f}\right)=\left\langle  e^{-\Delta S} \right\rangle
$ (I dropped the subscript $g$ since, from now on, I will be
interested in the torus only).

In general, one wishes to compute exactly the string amplitudes
(\eun), i.e. to all orders in
$\alpha '$. This is possible provided $\Delta S$ is a conformal
deformation, which requires the vertices appearing in (\edeux) to
form a set of truly marginal operators.
Moreover,
$Z\left({\bf f}\right)$
should be computed for finite backgrounds since the back-reaction of
gravity,
which one would like to take properly into account, appears generally
in the next-to-leading order in
$f_j$. Situations where all these requirements are satisfied appear
when, in string lattice compactifications,
$V_j(z,{\bar z})$ generate lattice Lorentz boosts. To clarify this
issue I will concentrate on the case where the deformation is
generated by
$V(z,{\bar z})=f_1 \, J(z)\,\overline{J}_1({\bar z})+
f_2 \, J(z)\, \overline{J}_2({\bar z})$.
Here $J(z)$ and $\overline{J}_{1,2}({\bar z})$
are $(1,0)$ and $(0,1)$
currents\footnote{$^{\,\ast}$}{\sevenrm The normalization for these
currents is such that
$\scriptstyle J(z)\,J(w)= {1\over(z-w)^2}+\cdots$
and similarly for the right-moving ones. If they are elements of an
affine Lie
algebra at level 1, this normalization implies that the higher root
has
$\scriptstyle \psi^2 =2$.}; moreover, I assume that
$\overline{J}_{1}({\bar z})$ and
$\overline{J}_{2}({\bar z})$ commute, and thus
$V(z,{\bar z})$ is an exactly marginal operator.
Their lattice momenta are $Q$ and ${\overline{Q}}_{1, 2}$
respectively. In terms of these charges, the undeformed one-loop
partition function can be written in the Hamiltonian approach as
follows:
$$
Z=\Tr \exp \Big(
-2\pi \Im \tau \left( L_0 + \overline{L}_0\right)
+2\pi i \Re \tau \left( L_0 - \overline{L}_0\right)
\Big)          \, ,                           \equn\put\etrois
$$
where $\tau$ is the genus-1 modular parameter and
$$
L_0={1\over 2}Q^2+\cdots  \, ,  \;
\overline{L}_0={1\over 2} \overline{Q}_1^{\, 2} +
{1\over 2} {\overline{Q}}_2^{\, 2}
+\cdots            \, ;                             \equn\put\equatre
$$
the dots stand for charges other than $Q$ and
${\overline{Q}}_{1, 2}$. The perturbation generated by
$
\Delta S=-{1\over 16\pi ^3}\int  \! 2i\, dz\,d{\bar z} \, V(z,{\bar
z})
$ turns out to be a Lorentz boost that can be easily implemented in
the Hamiltonian description (\etrois) and (\equatre). The action of
this boost on the charges is
$$
\pmatrix{Q_{\hphantom{.}}'\cr \overline{Q}_{1}' \cr \overline{Q}_{2}'
\cr }=
\pmatrix{\cosh \phi &\sinh \phi &0 \cr
\sinh \phi &\cosh \phi &0 \cr 0&0&1 \cr }
\pmatrix{1 &0&0 \cr 0 &{\hphantom{-}}\cos \theta&\sin\theta\cr 0
&-\sin\theta&\cos \theta \cr}
\pmatrix{Q_{\phantom{1}} \cr \overline{Q}_{1} \cr \overline{Q}_{2}
\cr }
                 \, ,                              \equn\put\ecinq
$$
leading to
$$
\eqalignno{
\Delta \left( L_0 + \overline{L}_0\right)
=&{\sqrt{1+{f_1^2 + f_2^2\over 64\pi^4}}-1\over 2}
\left(Q^2 + {\left(f_1 \, {\overline{Q}}_1 + f_2 \, {\overline{Q}}_2
\right)^2\over f_1^2 + f_2^2}\right)\cr
&\, + {f_1 \over 8\pi ^2}\, Q\, {\overline{Q}}_1
+{f_2 \over 8\pi ^2}\, Q\, {\overline{Q}}_2
\, ,
&\equnal\put\esix\cr }
$$
while
$ L_0 - \overline{L}_0$ remains invariant. To obtain eq. (\esix) I
identified the constant backgrounds
$f_1$ and $f_2$ with
$8 \pi^2 \sinh 2\phi \cos \theta$ and
$8 \pi^2 \sinh 2\phi \sin \theta$, respectively.
The deformed partition function now reads
$$
Z\left({\bf f}\right)=\Tr \exp \Big(
-2\pi \Im \tau \left( L_0 + \overline{L}_0 +
\Delta \left( L_0 + \overline{L}_0\right)\right)
+2\pi i \Re \tau \left( L_0 - \overline{L}_0\right)
\Big)
\, ,
\equn\put\esept
$$
which is exact for any finite value of $f_j$, and contains the
gravity back-reaction generated by these non-vanishing background
fields. Derivatives with respect to $f_j$ allow one to compute any
correlation function, provided the insertion of the charges $Q$ and
${\overline{Q}}_{1, 2}$ into the trace is tractable, which turns out
to be generally the case as will be seen in the sequel.

The method described above is general and can be applied successfully
to many situations. In the following, I will stick to a particular
string vacuum, namely the symmetric
$Z_2 \times Z_2$
orbifold in the heterotic construction. This model has $N=1$
space-time supersymmetry and an
$E_8 \times E_6 \times U(1)^2$
gauge group, which is promoted to
$E_8 \times E_6 \times SU(2)\times U(1)$
or even $E_8 \times E_6 \times SU(3)$
at some special values of the moduli describing  the internal
manifold. The latter is a six-dimensional torus parametrized by three
pairs of complex moduli:
$T_i\, ,\; U_i\, ,\; i=1,2,3$.
I choose this model because it has the advantage of reproducing
generic features, although it is quite simple. An interesting
property is, for instance, the absence of ($N=1$)-sector
contributions to the beta-function coefficients, which avoids the
related holomorphic anomalies.

The partition function for the
$Z_2 \times Z_2$
orbifold heterotic compactification reads
$$
\eqalignno{
Z(\tau, \bar \tau )=&{1 \over \Im \tau (\eta \bar\eta)^2}\,
 {1\over 2}\sum_{a,b}(-)^{a+b+ab}\,
 {\vartheta{a\atopwithdelims[]b}\over \eta} \cr
&\, {1\over 4}\sum_{h_j,g_j}
 {\vartheta{a+h_1\atopwithdelims[]b+g_1}\over \eta}
{\vartheta{a+h_2\atopwithdelims[]b+g_2}\over \eta}
 {\vartheta{a+h_3\atopwithdelims[]b+g_3}\over \eta}\,
 Z_{2,2}{h_1\atopwithdelims[]g_1}\,
 Z_{2,2}{h_2\atopwithdelims[]g_2}\,
 Z_{2,2}{h_3\atopwithdelims[]g_3}\cr
&\, {1\over 2}\sum_{\bar a,\bar b}
 {\bar\vartheta{\bar a+h_1\atopwithdelims[]\bar b+g_1}\over \bar\eta}
{\bar\vartheta{\bar a+h_2\atopwithdelims[]\bar b+g_2}\over \bar\eta}
 {\bar\vartheta{\bar a+h_3\atopwithdelims[]\bar b+g_3}\over \bar\eta}
 \left({\bar\vartheta{\bar a\atopwithdelims[]\bar b\vphantom{g}}\over
 \bar\eta}\right)^5
 {1\over 2}\sum_{\bar c,\bar d}
 \left({\bar\vartheta{\bar c\atopwithdelims[]\bar d\vphantom{g}}\over
 \bar\eta}\right)^8\, ,                     &\equnal\put\ehuit\cr}
$$
where the summation over the orbifold sectors is subject to the
condition
$h_1 +h_2 +h_3=0$
and similarly for the $g_j$'s. The first line of eq. (\ehuit)
corresponds to the transverse space-time coordinates together with
their left-moving supersymmetric partners; the internal bosonic
contribution together with the corresponding left-moving fermionic
one are given in the second line; the last line is associated with
the 32 right-moving fermions that generate the gauge group as well
as, together with the left-moving fermions, the internal
$(2,2)$  superconformal symmetry. The internal bosons
$X^4,\ldots,X^9$
are compactified on three two-tori. In the following sections, I will
use their complex combinations
$$
\eqalign{
\hphantom{^-}Y^{1}={X^4 +iX^5 \over \sqrt{2}}\, &, \;
\hphantom{^-}Y^{2}={X^6 +iX^7 \over \sqrt{2}}\, , \;
\hphantom{^-}Y^{3}={X^8 +iX^9 \over \sqrt{2}}\, , \cr
Y^{-1}={X^4 -iX^5 \over \sqrt{2}}\, &, \;
Y^{-2}={X^6 -iX^7 \over \sqrt{2}}\, , \;
Y^{-3}={X^8 -iX^9 \over \sqrt{2}}\, , \cr
}                                               \equn\put\eneuf
$$
together with their fermionic partners
$$
\eqalign{
\hphantom{^-}\Psi^{1}={\psi^4 +i\psi^5 \over \sqrt{2}}\, &, \;
\hphantom{^-}\Psi^{2}={\psi^6 +i\psi^7 \over \sqrt{2}}\, , \;
\hphantom{^-}\Psi^{3}={\psi^8 +i\psi^9 \over \sqrt{2}}\, , \cr
\Psi^{-1}={\psi^4 -i\psi^5 \over \sqrt{2}}\, &, \;
\Psi^{-2}={\psi^6 -i\psi^7 \over \sqrt{2}}\, , \;
\Psi^{-3}={\psi^8 -i\psi^9 \over \sqrt{2}}\, . \cr
}                                               \equn\put\edix
$$
In our conventions\footnote{$^{\,\dagger}$}{\sevenrm
Here, when the real parts of the moduli vanish, the imaginary parts
read
$\scriptstyle \Im T_j^{\phantom 1}=R_j^1\,R_j^2$
and
$\scriptstyle \Im U_j^{\phantom 1}=R_j^1 / R_j^2$, where
$\scriptstyle R_j^1$
and $\scriptstyle R_j^2$ are the compactification radii for the
$\scriptstyle j$th plane. Notice also that $\scriptstyle \alpha '
=1$.}
the two-torus partition function with periodic boundary conditions
for both compactified bosons reads
$$
\eqalignno{
Z_{2,2}{0\atopwithdelims[]0}\left(T_j,U_j,
\overline{T}_j,\overline{U}_j\right)
={1\over(\eta \bar \eta)^2}
&\Gamma_{2,2}\left(T_j,U_j,\overline{T}_j,\overline{U}_j\right) \cr
={1\over(\eta \bar \eta)^2}&\sum_{{\bf m,n}}
q^{\left| P^L_j \right|^2}{\bar q}^{\left| P^R_j\right|^2}
\, ,
&\equnal\put\eonze\cr}
$$
where
$$
\eqalignno{
\pmatrix{P^L_j \cr P^R_j}=&{1 \over 2i \sqrt{\Im T_j \Im U_j}}\times
\cr
&\times
\left\{
 m_1
\pmatrix{U^{\vphantom L}_j \cr U^{\vphantom R}_j}
-m_2
\pmatrix{1^{\vphantom L}_{\vphantom j} \cr
1^{\vphantom R}_{\vphantom j}}
+n_1
\pmatrix{T^{\vphantom L}_j \cr \overline{T}^{\vphantom .}_j}
+n_2
\pmatrix{T^{\vphantom L}_j \,U^{\vphantom L}_j\cr
\overline{T}^{\vphantom .}_j \, U^{\vphantom L}_j}
\right\}
&\equnal\put\edouze\cr}
$$
are the momenta associated with the left-moving
$i{\partial}Y^j$
and right-moving
$i{\bar \partial}Y^j$ currents. Similarly
$\pmatrix{P^{L\ast}_j  &P^{R\ast}_j}$ correspond to
$i{\partial}Y^{-j}$ and
$i{\bar \partial}Y^{-j}$.
On the other hand, when the bosons are twisted,
$$
Z_{2,2}{h_j \atopwithdelims[] g_j}=
{4 \eta {\bar \eta} \over \sqrt{
\vartheta {1+h_j \atopwithdelims[] 1+g_j}\,
\vartheta {1-h_j \atopwithdelims[] 1-g_j}\,
{\bar \vartheta} {1+h_j \atopwithdelims[] 1+g_j}\,
{\bar \vartheta} {1-h_j \atopwithdelims[] 1-g_j}
}}\, ,
                                              \equn\put\etreize
$$
which is moduli-independent.

The perturbative formulation of string theory does not allow one to
go easily off-shell in the string amplitudes. This does not affect so
much the sphere computations, but once one goes to the loops, the
presence of massless degrees of freedom leads to infra-red
divergences which, on the torus, appear at large values of $\Im
\tau$. In a recent article [\cinq] a
method for
regulating consistently these divergences has been developed. It
consists of replacing the four-dimensional flat space-time with a
more general $\sigma$-model preserving gauge symmetries,
supersymmetry and modular invariance and with a curvature that
induces a mass gap acting as an infra-red regulator. Among other
requirements, this $\sigma$-model must be $N=4$ superconformal in
order to be able to accommodate up to two space-time
supersymmetries\footnote{$^{\,\ast}$}{\sevenrm
In the above heterotic string theory, when the space-time
$\scriptstyle \sigma $-model is coupled with the internal theory, the
superconformal invariance reduces to $\scriptstyle N=2$.}.
This leads to several candidates [\dix] among which the simplest for
the above purposes is a $(1,0)$ supersymmetric version of
$W_k^{(4)}=U(1)_Q^{\vphantom )} \times SU(2)_k^{\vphantom )}$, with
$Q=\sqrt{1 \over k+2}$ a background charge for the time coordinate
chosen such that ${\hat c}_{\rm space-time}=4$. In practice (see
[\cinq, \six] for the details) this modification to the flat space
accounts for an extra factor in the partition function (eq. (\ehuit))
corresponding to the (suitably normalized) $SU(2)_k$ partition
function\footnote{$^{\,\dagger}$} {\sevenrm
Actually a $\scriptstyle Z_2 $-orbifold version of the $\scriptstyle
SU(2)_k$ WZW model
(or, put differently, a $\scriptstyle SO(3)_{k/ 2}$ model)
is necessary to guarantee the exact matching of
all degrees of freedom at the limit $\scriptstyle k\to \infty$, when
flat space-time is reached.}
$$
\Gamma(\mu)=-2\mu ^2 \sqrt{\Im \tau}{\partial \over \partial \mu}
\big[\Gamma_{1,1}(\mu)
-\Gamma_{1,1}(2\mu)
\big]
\bigg\vert_{\mu = {1\over \sqrt{k+2}}}\, ,
                                       \equn\put\equatorze
$$
with
$$
\Gamma_{1,1}(\mu)= \sum_{m,n}e^{{i\pi\tau\over2}\left(m\mu+{n\over
\mu}\right)^2}
e^{-{i\pi\overline{\tau}\over 2}\left(m\mu-{n\over \mu}\right)^2}\, .
                                       \equn\put\equinze
$$
This extra factor $\Gamma(\mu)$ ensures the convergence of the
integrals at large values of $\Im \tau$ by introducing a universal
mass gap
$\Delta m^2 = {\mu^2\over 2}$
with
$\mu = {1\over \sqrt{k+2}}$ (in $M_s\equiv {1\over
\sqrt{\alpha '}}$  units) to all (bosonic and fermionic) string
excitations. This infra-red regularization of the string (on-shell)
loop amplitudes vanishes when the flat-space limit is reached since
$\lim _{\mu \to 0} \Gamma(\mu)=1$.

Before going further in the application of the methods described so
far, let me comment on another consequence of the curvature in the
space-time sector of the model at hand. The introduction of a curved
background not only regulates the infra-red but makes it possible for
vertices such as the chromo-magnetic field, which are not
well-defined conformal operators on the flat space, to become truly
marginal on the $\sigma$-model version. This is precisely the feature
that allows the resolution of the $Z_2\times Z_2$ model in the
presence of a finite constant chromo-magnetic background, leading
therefore to the exact one-loop corrections for the gauge couplings
[\cinq]. More generally, an interesting question that one can address
is whether a suitable choice of space-time background is possible in
order to promote a given vertex to the rank of an exact conformal
operator.
\vskip 0.4cm
{\bf 3. $D$ auxiliary fields and the renormalization of the gauge
coupling}

I now turn to the computation of string amplitudes involving
auxiliary $D_{(\alpha)a}$ fields that belong to the vector multiplet;
here $\alpha$ labels the gauge-group factor $G_{\alpha}$ and $a$  is
the adjoint representation index. The (zero-momentum) vertex
operators for these fields are obtained by acting on the gauge-field
vertex in the minus-one-ghost picture with an appropriate form
[\onze]. The result is the operator
$$
V\left( D_{(\alpha)a} \right)={1\over \sqrt{3}}\left(
\Psi^{-1}\Psi^1 +
\Psi^{-2}\Psi^2 +
\Psi^{-3}\Psi^3 \right)
{\overline{J}_{(\alpha)a} \over \sqrt{k_{\alpha}}}\, ,
                                       \equn\put\eseize
$$
formally in the zero-ghost picture though it is not physical since it
does not survive the GSO projection. The bilinears $\Psi^{-j}\Psi^j$
are the internal helicity operators, one for each of the three
planes, and $Q^j$ the corresponding fermionic charges. The
left-moving factor of (\eseize) is the internal $N=2$ superconformal
current and $\overline{J}_{(\alpha)a} $ are elements of the affine
Lie algebra that realizes the gauge group; $\overline{Q}_{(\alpha)a}
$ are their lattice momenta. It deserves stressing here that these
results are generic for any symmetric orbifold $(2,2)$ heterotic
compactification; this is the reason why I introduced explicitly the
level $k_{\alpha}$ of the affine Lie
algebra\footnote{$^{\,\ast}$}{\sevenrm
I assume again that the highest root of the algebra has length
squared equal to 2. Therefore, the residue of the short-distance
leading singularities of bilinears in $\scriptstyle
{\overline{J}_{(\alpha)a} \over \sqrt{k_{\alpha}}}$ is 1. This agrees
with the normalizations of section 2.
Note, however, that in the effective field theory the usual
normalizations for the group algebra are such that
$\scriptstyle \psi ^2=1$, and one has to be careful when
identifying the effective renormalization constants with the
corresponding string amplitudes. This has been properly taken
into account throughout these notes.},
which is equal to 1 for all group factors in the $Z_2\times Z_2$
model.

Being the product of a left times a right current, the set of
operators $V\left( D_{(\alpha)a} \right)$ with index $a$
corresponding to the Cartan subalgebra generate an exactly integrable
$r$-dimensional deformation, where $r=\sum_{\alpha} r_{\alpha}$ is
the rank of the gauge group. The deformation is given by the
$SO(1,r)/SO(r)$ Lorentz boosts. To get a flavour of the effect of an
auxiliary-field background on the string, it is not necessary to look
at the most general deformation and I will actually restrict the
following analysis to the case of a single Cartan direction in the
group factor $G_{\alpha}$. The perturbation under consideration, at
the level of the action, is now
$$
\Delta S_{\alpha }=-{d_{\alpha}\over 16 \pi ^3}\int 2i\, dz \,
d\bar z \, V\left( D_{(\alpha)} \right)
                                       \equn\put\edixsept
$$
(I dropped the index $a$ by choosing a direction and there is no
summation over $\alpha$), and the deformed toroidal partition
function
$Z\left( d_{\alpha} \right)$
is given by eq.
(\esept)
with
$$
\eqalignno{\Delta \left( L_0 + \overline{L}_0\right)
=&{\sqrt{1+{d_{\alpha}^2 \over 64\pi^4}}-1\over 2}
\left( {\left(Q^1 +Q^2+Q^3 \right)^2_{\vphantom{.}}\over 3}+
{\overline{Q}_{(\alpha)}^2 \over k_{\alpha}}
\right)\cr
&\,+{d_{\alpha} \over 8\pi ^2}\, {Q^1 +Q^2+Q^3 \over\sqrt{
3}}\, {\overline{Q}_{(\alpha)}\over \sqrt{k_{\alpha}}}
\, .
                                       &\equnal\put\edixhuit\cr}
$$
This expression does not vanish as long as $d_{\alpha}\ne 0$ since a
constant $D$-field deformation ( (\eseize), (\edixsept)) breaks the
$(2,2)$ superconformal invariance (it preserves however the $(1,0)$
world-sheet supersymmetry).

Recalling that in the low-energy effective action the
vector-multiplet kinetic terms are of the form
$$
\sum_{\alpha ,a} {1\over
g_{\alpha}^2}
\left(
-{1\over 4}\, F_{(\alpha)a\mu \nu}^{\vphantom{\mu}}\,
F_{(\alpha)a}^{\phantom{(a)a}\mu \nu}
+{1\over 2}\, D_{(\alpha)a}^{\vphantom{\mu}}\,
D_{(\alpha)a}^{\vphantom{\mu}}
\right)
\,   ,
                                       \equn\put\edixneuf
$$
the one-loop string correction to $16\pi ^2 \over g_{\alpha}^2$ is
given by
$$
Z_{2,d_{\alpha}}=16\pi ^2 k_{\alpha}\int_{\cal F}
{d^2\tau \over (\Im \tau)^2}
{\partial^2 Z\left( d_{\alpha} \right)\over \partial d_{\alpha}^2}
\bigg\vert_{d_{\alpha}=0}
\,   .
                                       \equn\put\evingt
$$
Equations (\esept) and (\edixhuit) allow one to compute the above
derivative with the result:
$$
\eqalignno{
16\pi ^2 k_{\alpha}
{\partial^2 Z\over \partial d_{\alpha}^2}
\bigg\vert_{d_{\alpha}=0}=\Tr
\Biggl\{
&\exp \Big(-2\pi  \Im \tau\left(L_0+\overline{L}_0\right)
         +2\pi i\Re \tau\left(L_0-\overline{L}_0\right)\Big)
\cr
&(\Im \tau)^2 \Bigg(
{\left(Q^1 +Q^2+Q^3 \right)^2_{\vphantom{.}}\over
3}\overline{Q}_{(\alpha)}^2
\cr
&-{k_{\alpha} \over 4\pi\Im\tau}
\left[
{\left(Q^1 +Q^2+Q^3 \right)^2_{\vphantom{.}}\over 3}+
{\overline{Q}_{(\alpha)}^2 \over k_{\alpha}}
\right]\Bigg)
\Biggr\}\, .&\equnal\put\evingtetun\cr
}
$$
The last term of the post-exponential factor vanishes since the
insertion of $\overline{Q}_{(\alpha)}^2 $ does not affect the
supersymmetry properties of the left-moving sector. All I have to do
is therefore to compute the vacuum trace with the insertion of

$$
(\Im \tau)^2
{\left(Q^1 +Q^2+Q^3 \right)^2\over
3}\left(\overline{Q}_{(\alpha)}^2
-{k_{\alpha} \over 4\pi\Im\tau}\right)
\, .
                                       \equn\put\evingtdeux
$$
This insertion can be performed by noting that it is equivalent to
the action of a differential operator on the undeformed vacuum
amplitude. Indeed, $Q^j$ acts as ${1 \over 2\pi i}\left.{\partial
\over \partial v_j}\right\vert_{v_j=0}$ on the $i$th-plane
holomorphic $\vartheta $ function, while $\overline{Q}_{(\alpha)}^2 $
acts as ${i \over \pi}{\partial \over \partial \bar \tau}$ on the
appropriate subfactor of the 32 right-moving-fermion contribution.

I will now focus on the explicit computation of the above trace (eq.
(\evingtetun)) in the specific case of the $Z_2\times Z_2$ orbifold
model. The partition function for that model is given by (\ehuit),
multiplied by the extra factor (\equatorze) for the infra-red
regularization. One can use the generalized Jacobi identity (valid
under the condition $\sum_j h_j=\sum_j g_j=0$)
$$
\eqalignno{
{1\over 2}\sum_{a,b}
(-)^{a+b+ab}\,
\vartheta {a_{\vphantom j}\atopwithdelims[] b_{\vphantom
j}}\left(v_0\mid \tau\right)\, &\prod_{j=1}^{3}
\vartheta {a+h_j \atopwithdelims[] b+g_j}\left(v_j\mid
\tau\right)=\cr
=\vartheta {1_{\vphantom j}\atopwithdelims[] 1_{\vphantom
j}}\bigg({v_0-\sum_j v_j\over 2}\bigg\vert \tau\bigg)\,
&\prod_{j=1}^{3} \vartheta {1-h_j \atopwithdelims[] 1-g_j}
\bigg({v_0+\cdots -v_j+\cdots\over 2}\bigg\vert
\tau\bigg)
&\equnal\put\evingttrois\cr
}
$$
to recast (\ehuit) in a form where the insertion of  $\,{\left(Q^1
+Q^2+Q^3 \right)^2\over
3}\equiv -{1\over 12 \pi^2}\left. \left(
\sum_j{\partial \over \partial v_j}
\right)^2\right\vert_{{\bf v}=0}$
is more transparent.
Indeed, it becomes clear that the only non-zero contributions appear
when each of the two derivatives acts on a $\vartheta_1$. Hence,
only the $N=2$ sectors contribute (notice that for other orbifold
models such as the $Z_3$, there might exist non-vanishing
($N=1$)-sector contributions). Using the identities
${\vartheta_1'(0\mid \tau) \over 2\pi}=\eta^3(\tau)
={\vartheta_2^{\vphantom \prime}(0\mid \tau)\,
\vartheta_3^{\vphantom \prime}(0\mid \tau)\,
\vartheta_4^{\vphantom \prime}(0\mid \tau)\over 2}$,
one finally obtains
$$
Z_{2,d_{\alpha}}(\mu)=\int_{\cal F}
{d^2\tau \over \Im \tau}\,
\Gamma(\mu)\, \sum_j {\Gamma_{2,2}(j)
\over {\bar \eta}^{24}}
\left(\overline{Q}_{(\alpha)}^2
-{k_{\alpha} \over 4\pi\Im\tau}\right)\overline{\Omega }
                   \, ,                    \equn\put\evingtquatre
$$
where
$k_{\alpha}=1$,
$\Gamma_{2,2}(j)\equiv
\Gamma_{2,2}\left(T_j,U_j,\overline{T}_j,\overline{U}_j\right)$
is given by (\eonze), $\Gamma(\mu)$
is the infra-red regulating factor (see eqs.
(\equatorze) and  (\equinze)), and $\overline{\Omega
}=\overline{\Omega }_8  \, \overline{\Omega }_6$, with
$$
{\Omega }_8^{\phantom 8}=
{1\over 2}\sum_{a}{\vartheta}^8_{a }\
{\rm and}   \
{\Omega}_6^{\phantom 4}={1\over 4}
\left({\vartheta}_2^4 + {\vartheta}_3^4\right)
\left({\vartheta}_3^4 + {\vartheta}_4^4\right)
\left({\vartheta}_2^4 - {\vartheta}_4^4\right)
                                       \equn\put\evingtcinq
$$
(these are proportional to the Eisenstein modular-covariant functions
$E_2$ (or $G_2$) and $E_3$ (or $G_3$), respectively:
$\Omega _8 = E_2 ={G_2 \over 2\zeta(4)}$ and
$-2 \Omega _6 = E_3 ={G_3 \over 2\zeta(6)}$
[\douze]\footnote{$^{\,\dagger}$}{\sevenrm
Some authors (as in ref. [\quatorze]) use the notation
$\scriptstyle E_{2k} $
instead of
$\scriptstyle E_{k} $,
$\scriptstyle k\geq 2$.}).
Notice that the contribution to the partition function of the twisted
bosons cancels that of the twisted fermions. Although the result
(\evingtquatre) has been achieved in the framework of the $Z_2 \times
Z_2$ model, it actually applies to any $(2,2)$ symmetric orbifold
once the appropriate modifications are performed at the level of the
moduli-dependent function
$\sum_j \Gamma_{2,2}(j)$
as well as of the modular function $\overline{\Omega }$ [\thgen]. It
is interesting to observe that the radiative corrections
(\evingtquatre) include exactly the back-reaction of the
gravitationally coupled fields; this accounts for the term
${-k_{\alpha} \over 4\pi\Im\tau}$, which is universal and guarantees
modular invariance.

An important conclusion that the above analysis allows one to draw is
the following: the one-loop correction to $16\pi ^2 \over
g_{\alpha}^2$ calculated here (expression (\evingtquatre)) coincides
with the one that was first obtained in [\cinq] by considering string
amplitudes with two magnetic field insertions. As appears from the
low-energy field theory (see eq. (\edixneuf)), the one-loop
correction to $16\pi ^2 \over g_{\alpha}^2$ is indeed  expected to be
given by
$$
Z_{2,b_{\alpha}}=-16\pi ^2 k_{\alpha}\int_{\cal F}
{d^2\tau \over (\Im \tau)^2}
{\partial^2 Z\left( b_{\alpha} \right)\over \partial b_{\alpha}^2}
\bigg\vert_{b_{\alpha}=0}
                                       \equn\put\evingtsix
$$
provided the supersymmetry Ward identity that relates the
$F_{(\alpha)a}^{\phantom{(a)a}\mu \nu}$-
and $D_{(\alpha)a}^{\vphantom{\mu}}$-field
renormalization holds at the level of the fundamental theory. In eq.
(\evingtsix) $Z\left( b_{\alpha} \right)$ is the partition function
in the presence of a constant background chromo-magnetic field
$b_{\alpha}$ pointing for instance in the third space direction, and
in some Cartan direction of the group algebra. In flat space,
two-dimensional conformal invariance is broken by a constant magnetic
background because of gravity back-reaction. As already mentioned in
the previous section, however, in the $W_k^{(4)}$
$\sigma$-model, a constant magnetic field induces a conformal
deformation generated by
$$
\Delta S_{\alpha }=-{b_{\alpha}\over 16 \pi ^3}\int 2i\, dz \,
d\bar z \,{J^3 + i : \psi^1 \psi^2 :\over \sqrt{{k\over 2}+1}}\,
{\overline{J}_{(\alpha)}^{\rm \ any\ Cartan} \over \sqrt{k_{\alpha}}}
                                       \equn\put\evingtsept
$$
and leads to the result\footnote{$^{\,\ast}$}{\sevenrm
In order to take into account the finite volume of the $\scriptstyle
SU(2)_k$ manifold and recover the correct smooth flat-space limit at
$\scriptstyle k\to \infty$, the relation between the magnetic field
and the boost parameter is here $\scriptstyle b_{\alpha}={8\pi ^2
\sinh 2\phi\over \sqrt{{k/2}+1}}$.}:
$$
\eqalignno{
16\pi ^2 k_{\alpha}
{\partial^2 Z\over \partial b_{\alpha}^2}
\bigg\vert_{b_{\alpha}=0}= \Tr
\biggl\{
&\exp\Big(-2\pi  \Im \tau\left(L_0+\overline{L}_0\right)
         +2\pi i\Re \tau\left(L_0-\overline{L}_0\right)\Big)
\cr
&(\Im \tau)^2 \bigg(
(I+Q)^2_{\vphantom{/_/}}\,
\overline{Q}_{(\alpha)}^2
\cr
&-{k_{\alpha} \over 4\pi\Im\tau}
\left[
(I+Q)^2_{\vphantom{/_/}}+
\overline{Q}_{(\alpha)}^2 {k+2\over 2 k_{\alpha}}
\right]\bigg)
\biggr\}\,.                            &\equnal\put\evingthuit\cr
}
$$
Here $J^3$ is a $SO(3)_{k\over 2}$ current, $I$ is the corresponding
zero mode
and $Q$ are the fermionic charges. As a consequence of supersymmetry
properties, the insertion of $I^2$ and $\overline{Q}_{(\alpha)}^2$
vanish; that of $IQ$ could contribute in the presence of $N=1$
sectors but is actually zero because of space isotropy. Finally the
insertion of $Q^2$ amounts to the action of $-{1 \over 4\pi
^2}\left.{\partial ^2\over \partial v_0^2}\right\vert_{v_0=0}$ on the
holomorphic functions $\vartheta {a \atopwithdelims[] b}\left(v_0\mid
\tau\right)$, which, thanks to the generalized Jacobi identity, is
equivalent to the action of ${1 \over 12\pi
^2}\left.\left(\sum_j{\partial \over \partial
v_j}\right)^2\right\vert_{{\bf v}=0}$
 on $\prod_{j=1}^{3}
\vartheta {a+h_j \atopwithdelims[] b+g_j}\left(v_j\mid
\tau\right)$. This demonstrates that
$$
Z_{2,b_{\alpha}}=Z_{2,d_{\alpha}} \, ,
                                       \equn\put\evingtneuf
$$
which is the anticipated Ward identity, at the one-loop level.

Finally, I would like to emphasize that the one-loop corrections to
the gauge  couplings (\evingtquatre), as was extensively discussed in
[\cinq], are exact, i.e. that they contain rigorously both universal
and group-factor-dependent thresholds. The former were missing in the
pioneering works on string thresholds [\trois]. They actually
contain, besides the gravity back-reaction
term\footnote{$^{\,\dagger}$}{\sevenrm
This universal term was also found in [\sept] for the three-point
function of two gauge bosons and the modulus $\scriptstyle T$.}
${-k_{\alpha} \over 4\pi\Im\tau}$, other contributions originated
from the insertion of $\overline{Q}_{(\alpha)}^2$.
These contributions, which guarantee modular invariance and infra-red
finiteness, have been worked out for the symmetric $Z_2 \times Z_2$
orbifold
in [\treize ], where the effects of the (moduli-dependent) universal
thresholds on string unification are also analyzed; generalization to
a larger class of models can be found in [\thgen].
As far as the infra-red regularization is concerned, I should also
mention here that the regularized correlator $Z_{2,d_{\alpha}}(\mu)$
(eq. (\evingtquatre)) is related to the one-loop effective theory
running gauge coupling $g_{\alpha}(\mu)$ in a very simple way
[\treize]:
$$
{16\pi^2\over g_{\alpha}^2(\mu)} =
k_\alpha{16\pi^2\over g_s^2} +Z_{2,d_{\alpha}}(\mu)
- {\bf b}_\alpha(2\gamma+2)\, ,
                                       \equn\put\etrente
$$
where $g_s$ is the string coupling and ${\bf b}_{\alpha}$ the
beta-function coefficients. This expression holds in the
$\overline{DR}$ scheme and despite the presence of the
curvature-induced infra-red regulator $\Gamma (\mu)$ in
$Z_{2,d_{\alpha}}(\mu)$, the thresholds are infra-red cut-off
independent [\treize]. This shows how the above regularization
procedure allows one to avoid ambiguities.
\vskip 0.4cm
{\bf 4. One-loop corrections to the K\"ahler metric for moduli
fields}

The kinetic terms for the scalar fields and the corresponding
auxiliary $F$ fields appear in the low-energy effective action as
$$
K^{\bar j}_{\hphantom{j}i}\, \left(
\partial_{\mu}^{\vphantom{\bar j}}{\bar z}_j^{\vphantom{\bar j}}\,
\partial^{\mu}_{\vphantom{\bar j}}z^i_{\vphantom{\bar j}}
+
\overline{F}_j^{\vphantom{\bar j}}\, F^i_{\vphantom{\bar j}}
\right)   \, ,
                                     \equn\put\etrenteetun
$$
where $K^{\bar j}_{\hphantom{j}i}=
{\partial^2 K\over \partial {\bar z}_j\partial z^i}$
are the K\"ahler metric elements and $K(z,{\bar z})$ is the K\"ahler
potential. An important issue is the dependance of this potential on
the moduli fields, first because the K\"ahler geometry of the moduli
space is related to the Yukawa couplings of the matter fields, and
eventually because $K$ enters the scalar potential and could
influence supersymmetry-breaking mechanisms. I will focus here on the
particular case of the untwisted moduli
$T,\overline{T},U,\overline{U}$ that appear in symmetric orbifold
$(2,2)$ compactifications.

The moduli K\"ahler metric can be extracted from string scattering
amplitudes involving four complex moduli fields, by solving a
differential equation [\quinze]. Although such a procedure could be
helpful on the sphere, it becomes very complicated beyond tree level.
Of course, one might proceed to a direct determination of the
wave-function renormalization for moduli fields, by looking at
amplitudes that involve vertices associated with $\partial _{\mu} T$,
\dots \ Unfortunately it is difficult to perform exactly this kind of
computation in string theory, essentially because of its
first-quantized formulation. On the other hand, advocating space-time
supersymmetry, it can be argued that this is actually equivalent to
computing string amplitudes for two auxiliary fields, and this is
precisely what I will be analyzing in the following. This
straightforward method is an alternative to the previous indirect
determination of the one-loop K\"ahler metric performed by looking at
three-point functions of two moduli and one antisymmetric tensor
[\huit]. The actual calculation of the appropriate string amplitudes
will be realized here by switching on constant $F$-auxiliary-field
backgrounds, as I did in the previous section for the case of $D$
fields.

In order to proceed further, let me describe the vertices associated
with the moduli multiplets of the $Z_2 \times Z_2$ orbifold model. I
will concentrate on the third plane where, according to the
conventions (\eonze) and (\edouze), the correspondence is the
following:
$$\eqalign{
T_3^{\vphantom{X}}\, : \;
{-i\partial Y^{-3}\, {\bar \partial}Y^{3\hphantom{-}}\over \Im T_3}
\, &, \;\overline{T}_3\, : \;
{{\hphantom-}i\partial Y^{3\hphantom{-}}\, {\bar \partial}Y^{-3}\over
\Im T_3}
\, ,\cr
U_3^{\vphantom{X}}\, : \;
{{\hphantom-}i\partial Y^{-3}\, {\bar \partial}Y^{-3}\over \Im U_3}
\, &, \;\overline{U}_3\, : \;
{-i\partial Y^{3\hphantom{-}}\, {\bar \partial}Y^{3\hphantom{-}}\over
\Im U_3}
\, ,\cr}
                                       \equn\put\etrentedeux
$$
in the zero-ghost picture. The auxiliary fields are obtained by going
to the minus-one-ghost picture and acting with the top form (or its
complex conjugate) [\onze]. At this point it is convenient to
introduce the left-moving $SO(4)$ level-one current algebra generated
by bilinears of the first- and second-plane fermions
$\psi^4$, $\psi^5$, $\psi^6$, $\psi^7$:
$$
\eqalign{
J^x={:\psi^6 \psi^5 :
+ :\psi^7 \psi^4 :\over i\sqrt{2}}\, , \;
K^x={:\psi^6 \psi^5 :
- :\psi^7 \psi^4 :\over i\sqrt{2}}\, , \; \cr
J^y={:\psi^6 \psi^4 :
+ :\psi^5 \psi^7 :\over i\sqrt{2}}\, , \;
K^y={:\psi^6 \psi^4 :
- :\psi^5 \psi^7 :\over i\sqrt{2}}\, , \; \cr
J^z={:\psi^4 \psi^5 :
+ :\psi^6 \psi^7 :\over i\sqrt{2}}\, , \;
K^z={:\psi^4 \psi^5 :
- :\psi^6 \psi^7 :\over i\sqrt{2}}\, . \; \cr
}
                                       \equn\put\etrentetrois
$$
In this form the
$SU(2)\times SU(2)$
structure is manifest and the normalizations are such that for each
$SU(2)$ algebra the roots have length squared equal to 2.
With the above definitions the vertices for the auxiliary fields
read:
$$\eqalign{
F_{T_3}^{\vphantom{X}}\, : \;
{-
\big(\left(K^y - J^y\right)+i\left(K^y + J^y\right)\big)\, {\bar
\partial}Y^{3\hphantom{-}}\over 2i\Im T_3}
\, &, \cr
\overline{F}_{T_3}\, : \;
{-
\big(\left(K^y - J^y\right)-i\left(K^y + J^y\right)\big)\, {\bar
\partial}Y^{-3}\over 2i\Im T_3}
\, &, \cr
F_{U_3}^{\vphantom{X}}\, : \;
{{\hphantom-}
\big(\left(K^y - J^y\right)+i\left(K^y + J^y\right)\big)\, {\bar
\partial}Y^{-3}\over 2i\Im U_3}
\, &, \cr
\overline{F}_{U_3}\, : \;
{{\hphantom-}
\big(\left(K^y - J^y\right)-i\left(K^y + J^y\right)\big)\, {\bar
\partial}Y^{3\hphantom{-}}\over 2i\Im U_3}
\, &. \cr}
                                      \equn\put\etrentequatre
$$
Obviously, both left and right factors are Abelian currents;
therefore, switching on the corresponding backgrounds simultaneously
does not affect conformal invariance. Moreover if $f_{T_3}$, ${\bar
f}_{T_3}$,
$f_{U_3}$, ${\bar f}_{U_3}$
are these constant backgrounds, the one-loop corrections to the
K\"ahler metric read\footnote{$^{\,\ast}$}{\sevenrm
Remember that $\scriptstyle K^{\bar j}_{\hphantom{j}i}=
 K^{\bar j\, (0)}_{\hphantom{j}i}  +g_s^2\,
 K^{\bar j\, (1)}_{\hphantom{j}i} +\cdots
$ with $\scriptstyle K^{\overline{T}_3\, (0)}_{\hphantom{T_3}T_3}=
{1\over 4(\Im T_3)^2}\, ,\ldots  $
Notice also that operators such as those appearing in
(\etrentequatre) do not commute with the superconformal current.
Therefore, the above method becomes questionable when applied to
higher-order correlators or at higher genus.}
$$
K^{\overline{T}_3\, (1)}_{\hphantom{T_3}T_3}=
\int_{\cal F}
{d^2\tau \over (\Im \tau)^2}
{\partial^2 Z({\bf f})\over
\partial {\bar f}_{T_3}
\partial f_{T_3}}
\bigg\vert_{{\bf f}=0}
\, ,\;K^{\overline{T}_3\, (1)}_{\hphantom{T_3}U_3}=
\int_{\cal F}
{d^2\tau \over (\Im \tau)^2}
{\partial^2 Z({\bf f})\over
\partial {\bar f}_{T_3}
\partial f_{U_3}}
\bigg\vert_{{\bf f}=0}\, ,
                               \equn\put\etrentecinq
$$
and similarly for the other components.

There are a few observations that one can make in order to simplify
the computation of quantities such as $\left.{\partial^2 Z({\bf
f})\over
\partial {\bar f}_{T_3}
\partial f_{T_3}}
\right\vert_{{\bf f}=0}
$, ${\partial^2 Z({\bf f})\over
\partial {\bar f}_{T_3}
\partial f_{U_3}}
\Big\vert_{{\bf f}=0}
$, \dots \
First, the presence of the $SU(2) \times SU(2)$ symmetry allows a
rotation of $J^y$ and $K^y$  onto $J^z$ and $K^z$. The zero modes of
the latter are the combinations
$-{Q^1+Q^2\over \sqrt{2}}$ and $-{Q^1-Q^2\over \sqrt{2}}$
of the internal helicity operator charges introduced in the previous
section. When inserted in the vacuum trace, these charges act as
${i\over 2\pi \sqrt{2}}\left(
{\partial \over \partial v_1}+{\partial \over \partial v_2}\right)$
and
${i\over 2\pi \sqrt{2}}\left(
{\partial \over \partial v_1}-{\partial \over \partial v_2}\right)$
on the holomorphic $\vartheta$ functions of the first and second
plane. Secondly, instead of (\etrentequatre) it is more convenient to
choose the following basis of real (commuting) $(1, 1)$ operators:
$$
K^y\, i{\bar \partial}X^8\, ,\;
J^y\, i{\bar \partial}X^8\, ,\;
K^y\, i{\bar \partial}X^9\, ,\;
J^y\, i{\bar \partial}X^9\, ,\;
                               \equn\put\etrentesix
$$
and denote $\alpha$, $\beta$, $\gamma$, $\delta$ the corresponding
(real) backgrounds. A straightforward calculation shows that
$$
\eqalignno{
{\partial^2 \over
\partial {\bar f}_{T_3}
\partial f_{T_3}}={1\over 4 \left( \Im T_3\right)^2}
\bigg(
&-2{\partial ^2\over\partial \alpha\partial \delta}
+2{\partial ^2\over\partial \beta\partial \gamma}\cr
&\,+{\partial ^2\over\partial \alpha^2 }
+{\partial ^2\over\partial \beta^2 }
+{\partial ^2\over\partial \gamma^2 }
+{\partial ^2\over\partial \delta^2 }
\bigg)
\, ,
                               &\equnal\put\etrentesept\cr}
$$
$$
\eqalignno{
{\partial^2 \over
\partial {\bar f}_{T_3}
\partial f_{U_3}}={-1\over 4 \Im T_3 \Im U_3}
\bigg(
&-2i{\partial ^2\over\partial \alpha\partial \gamma}
-2i{\partial ^2\over\partial \beta\partial \delta}\cr
&\, +{\partial ^2\over\partial \alpha^2 }
+{\partial ^2\over\partial \beta^2 }
-{\partial ^2\over\partial \gamma^2 }
-{\partial ^2\over\partial \delta^2 }
\bigg)
\, ,
                               &\equnal\put\etrentehuit\cr}
$$
and similarly for the others. As far as the left-moving sector is
concerned\footnote{$^{\,\dagger}$}{\sevenrm
Insertions of operators acting exclusively on the right-moving
sectors vanish identically because of supersymmetry.},
derivatives such as ${\partial ^2\over\partial \alpha^2 }$,
${\partial ^2\over\partial \gamma^2 }$ or
${\partial ^2\over\partial \alpha\partial \gamma}$
will involve, according to the above argument, insertions of
$\left(Q^1 - Q^2\right)^2$ into the vacuum trace. Thanks to the
Jacobi identity (\evingttrois), this insertion vanishes. On the other
hand, the derivatives ${\partial ^2\over\partial \alpha\partial
\delta}$ and
${\partial ^2\over\partial \beta\partial \gamma}$ will lead to
insertions of $\left(Q^1 - Q^2\right)\left(Q^1 + Q^2\right)$. In
general, these do receive contributions from the $N=1$ sectors and it
is not difficult to see that those contributions are equal for both
${\partial ^2\over\partial \alpha\partial \delta}$ and ${\partial
^2\over\partial \beta\partial \gamma}$, and thus cancel in
(\etrentesept)
(this is of course irrelevant in the $Z_2\times Z_2$ orbifold where
there are no $N=1$ sectors). Hence, one concludes that
$$
{\partial^2 Z({\bf f})\over
\partial {\bar f}_{T_3}
\partial f_{T_3}}
\bigg\vert_{{\bf f}=0}={1\over 4 \left( \Im T_3\right)^2}
\left(
{\partial ^2\over\partial \beta^2 }+
{\partial ^2\over\partial \delta^2 }
\right) Z(\beta,\delta)\bigg\vert_{\beta=\delta=0}\, ,
                               \equn\put\etrenteneuf
$$
and
$$
{\partial^2 Z({\bf f})\over
\partial {\bar f}_{T_3}
\partial f_{U_3}}
\bigg\vert_{{\bf f}=0}={-1\over 4 \Im T_3 \Im U_3}
\left(
{\partial ^2\over\partial \beta^2 }-
{\partial ^2\over\partial \delta^2 }-2i
{\partial ^2\over\partial \beta\partial \delta}
\right)Z(\beta,\delta)\bigg\vert_{\beta=\delta=0}\, .
                               \equn\put\equarante
$$

Equations (\etrenteneuf) and (\equarante) show that, for the issue of
the one-loop corrections to the K\"ahler metric, it is not necessary
to analyze the most general deformation: only $Z(\beta,\delta)$ is
relevant. Such a deformation has been considered in section 2 for a
generic case (eqs.~(\esix) and (\esept)) and those results
can immediately be applied to (\etrenteneuf) and (\equarante):
$$
\eqalignno{
{\partial^2 Z({\bf f})\over
\partial {\bar f}_{T_3}
\partial f_{T_3}}
\bigg\vert_{{\bf f}=0}={(\Im \tau)^2\over 64\pi^2 \left( \Im
T_3\right)^2}
\Tr
\bigg\{
\left(Q^1 +Q^2\right)^2
\left[
\left\vert p^R_3\right\vert^2
-{1\over 4\pi\Im\tau}
\right]&\cr
\exp\Big(-2\pi  \Im \tau\left(L_0+\overline{L}_0\right)
         +2\pi i\Re \tau\left(L_0-\overline{L}_0\right)\Big)
&\bigg\} \, ,
                                       &\equnal\put\equaranteetun\cr}
$$
$$
\eqalignno{
{\partial^2 Z({\bf f})\over
\partial {\bar f}_{T_3}
\partial f_{U_3}}
\bigg\vert_{{\bf f}=0}={-(\Im \tau)^2\over 64 \pi ^2 \Im T_3 \Im U_3}
&\Tr
\biggl\{
\left(Q^1 +Q^2\right)^2
\left(
p^{R\ast}_3
\right)^2\cr
\exp\Big(-2\pi  \Im \tau&\left(L_0+\overline{L}_0\right)
         +2\pi i\Re \tau\left(L_0-\overline{L}_0\right)\Big)
\biggr\}\, .
                                       &\equnal\put\equarantedeux\cr}
$$
The evaluation of the traces is performed along the same lines of
thought as in the $D$-field calculation. Now, the only non-vanishing
contribution comes from the $N=2$ sectors with twist
$\left(h_3,g_3\right)=(0,0)$. After some
algebra, using in particular the fact that the insertion of
$\left\vert p^R_3\right\vert^2$ amounts to the operator ${i \over
2\pi}{\partial \over \partial \bar \tau}$ acting on the solitonic
contribution of the third two-torus, one obtains
$$
K^{\overline{T}_3\, (1)}_{\hphantom{T_3}T_3}=
{-1\over 16 \pi^2 \left( \Im T_3\right)^2}
\int_{\cal F}
{d^2\tau \over (\Im \tau)^2}
\,
\Gamma(\mu)\,{\overline{\Omega }
\over {\bar \eta}^{24}}\,
{i\over 2 \pi}
\, {\partial \over \partial \bar \tau}
\Big(
\Im \tau\,
\Gamma_{2,2}(3)
\Big)
\, ,
                                      \equn\put\equarantetrois
$$
$$
K^{\overline{T}_3\, (1)}_{\hphantom{T_3}U_3}=
{1\over 16 \pi ^2 \Im T_3 \Im U_3}
\int_{\cal F}
{d^2\tau \over \Im \tau}
\,
\Gamma(\mu)\,{\overline{\Omega }
\over {\bar \eta}^{24}}
\,
\sum_{\bf m,n}
\left(
p^{R\ast}_3
\right)^2q^{\left| P^L_3 \right|^2}{\bar q}^{\left| P^R_3\right|^2}
\, ,
                                       \equn\put\equarantequatre
$$
and similarly for the other elements.

The above results deserve a few comments.
The computation I presented here is exact, that is it holds to all
orders in $\alpha'$ and takes properly into account the back-reaction
of gravity (terms proportional to ${1\over 4\pi\Im\tau}$ in
(\equaranteetun) or (\equarantetrois)). The integrands of
(\equarantetrois) and (\equarantequatre) are $\tau$-modular-invariant
functions, which are well behaved at large values of $\Im \tau$.
Therefore,  one can remove the infra-red regulator by taking the
limit $\mu \to 0$ in the corresponding integrals. By using identities
such as
$$
{1 \over (\Im T)^2}\,
{\partial ^2\over \partial \bar \tau \partial \tau}
\Big(\Im  \tau \,
\Gamma_{2,2}
\left(T,U,\overline{T},\overline{U}\right)\Big)
={1\over \Im \tau }\,
{\partial ^2\over \partial \overline{T} \partial T}
\Gamma_{2,2}
\left(T,U,\overline{T},\overline{U}\right)
                                       \equn\put\equarantecinq
$$
and integrating by parts when appropriate, it is then easy to show
that
$$
K^{\overline{T}_3\, (1)}_{\hphantom{T_3}T_3}=
{\partial ^2 K^{(1)}
\over \partial \overline{T}_3 \partial T_3}
\, ,\;
K^{\overline{T}_3\, (1)}_{\hphantom{T_3}U_3}=
{\partial ^2 K^{(1)}
\over \partial \overline{T}_3 \partial U_3}
\, , \ldots  ,
                                       \equn\put\equarantesix
$$
where the real function
$$
K^{(1)}\left({\bf T},{\bf U},\overline{\bf T},\overline{\bf
U}\right)={1\over 4\pi^2}\,
Y\left({\bf T},{\bf U},\overline{\bf T},\overline{\bf U}\right) +
\kappa\big({\bf T},{\bf U}\big) +
{\bar\kappa}\left(\overline{\bf T},\overline{\bf U}\right)
                                       \equn\put\equarantesept
$$
is the one-loop correction to the K\"ahler potential. The first term
in (\equarantesept) is, as expected from supersymmetry, proportional
to the universal part of the one-loop thresholds of the gauge
couplings [\treize]:
$$
\eqalignno{
\!\!\! Y\left({\bf T},{\bf U},\overline{\bf T},\overline{\bf
U}\right)&=
{1\over 6}\sum_{i}
\log\big\vert j\left(T_i\right)-j\left(U_i\right)\big\vert
\cr &\;+
\int_{\cal F}{d^2\tau \over \Im \tau}
\,\sum_{i} \Gamma_{2,2}(i) \,
\left(-2\left[
{\partial \log {\bar \eta}\over \partial \log {\bar q}}\,
- {1\over 8\pi \Im\tau}\right]
{\overline{\Omega}\over \overline{\eta}^{24}}
+11
\right)
\, .
                                      &\equnal\put\equarantehuit\cr}
$$
This expression is invariant under duality transformations on
$T_i,U_i$ and finite everywhere in the moduli space despite the
logarithmic divergences that both terms suffer around enhanced
symmetry lines. On the other hand, $\kappa({\bf T},{\bf U})$ is an
analytic function of the moduli, which is irrelevant in
(\equarantesix) but plays an important role for the duality
covariance of $K^{(1)}$. Finally, it is interesting to observe that
due to the identity
$$
{\Omega\over \eta ^{24}}={j-j(i)\over 2}
\left({\partial \log j\over \partial \log q}\right)^{-1}\, ,
                                       \equn\put\equaranteneuf
$$
the corrections to the K\"ahler metric I determined for the specific
$N=1$ model turn out to be identical to those obtained in [\seize]
for a class of $N=2$ string vacua.
Expression (\equarantehuit), and consequently expression
(\equarantesept), can be further generalized to a larger class of
$N=2$ four-dimensional theories, namely those that come from
toroidal compactification of generic six-dimensional $N=1$ ground
states. These
theories always possess a universal two-torus. Therefore, advocating
$\tau $-modular invariance and infra-red finiteness all over the
moduli space, one can draw the conclusion that for these models the
universal thresholds are proportional to the contribution of a
single plane in the
$Z_2 \times Z_2$ orbifold. The coefficient can be related via the
gravitational anomaly to the quantity $N_v - N_h$, where $N_v$
and $N_h$ are the number of massless vector multiplets and the
number of massless hypermultiplets, respectively. Taking into account
that for this class of ground states the cancellation
of anomalies originated from six dimensions implies that
$N_v - N_h$ is a universal constant
[\sch], the net result reads [\thgen]:
$$
\eqalignno{
\!\!\!\!\!\!\!\!\!\!\!\!
Y\left(T,U,\overline{T},\overline{U}\right)&=
{1\over 3}
\log\big\vert j(T)-j(U)\big\vert
\cr
&\; +2\int_{\cal F}{d^2\tau \over \Im \tau}
\,\Gamma_{2,2}\left(T,U,\overline{T},\overline{U}\right)
\left(-2\left[
{\partial \log {\bar \eta}\over \partial \log {\bar q}}\,
- {1\over 8\pi \Im\tau}\right]
{\overline{\Omega}\over \overline{\eta}^{24}}
+11
\right)\,  .
                                      &\equnal\put\ecinquante\cr}
$$
In the presence of local $N=2$ space-time supersymmetry, the K\"ahler
potential is described in terms of an analytic function, the
prepotential. The moduli dependance of the latter, as well as its
properties under modular transformations, is non-trivial and has
attracted much attention in the rapidly growing subject of dualities
[\seize  --\quatorze]. This is another motivation for a careful study
of the one-loop K\"ahler potential.
\vskip 0.4cm
{\bf 5. Conclusions}

By applying the background  field method to situations where the
corresponding deformations are truly conformal, it has been possible
to compute exact one-loop string amplitudes for both vector $D$ and
chiral $F$ fields. The former lead to a supersymmetry Ward identity
relating amplitudes of various members of the vector
supermultiplet\footnote{$^{\,\ast}$}{\sevenrm
A similar Ward identity would have been reached by calculating
$\scriptstyle H$-field string amplitudes and comparing them with
$\scriptstyle R$-tensor insertions: they both lead to the
gravitational coupling corrections
(for a discussion see [\thgen]).};
in the framework of strings, this is a simple consequence of the
Jacobi identity between $\vartheta$ functions. The $F$-field
insertions, on the other hand, give the corrections to the K\"ahler
metric for the moduli fields, which turn out to contain information
about the K\"ahler potential.

The one-loop corrections to the K\"ahler potential are determined by
the universal thresholds of the gauge couplings as well as by the
real part of an analytic function $\kappa({\bf T},{\bf U})$. An
interesting open problem is the exact determination, i.e. up to a
degree-two polynomial, of this analytic function. In fact, the
knowledge of the K\"ahler potential as a function of the moduli is
useful (\romannumeral1)
for phenomenological purposes since it is related to the Yukawa
couplings and enters the scalar potential where it could play a role
when, in case of supersymmetry breaking, the moduli are no longer
flat
directions;
(\romannumeral2)~because,
for $N=2$ supersymmetric vacua, it is related to the prepotential.
Moreover, as I explained at the end of the previous section, although
the calculation I presented here strictly holds for the $Z_2 \times
Z_2$ orbifold, the results seem to have a wider application either in
the case of some $N=2$ superstring compactifications, or for the
($N=2$)-sector contributions of other $N = 1$ string vacua.

Finally, it can be observed that amplitudes involving $F$-auxiliary
fields associated with the moduli do not suffer from infra-red
divergences. However, it is clear from the $D$-field computation that
this is not a generic feature, and that for scalar fields other than
the moduli, one generally needs to keep $\mu$ finite and treat these
infra-red divergences as is done for the gauge couplings [\cinq,
\treize]. This would happen in particular in the case of charged
(twisted or untwisted) chiral multiplets for which it would be
interesting to compute the K\"ahler metric corrections and their
effects on the Yukawa couplings [\vingt].
\vskip 0.4cm
\vfil
\centerline{\bf Acknowledgements}

These notes are based on work done in collaboration with
E.~Kiritsis and C.~Kounnas. I would like also to thank J. Rizos for
stimulating  discussions. This work was  supported in part by
EEC contracts CHRX-CT93-0340  and SC1-CT92-0792.
\vskip 0.4cm
{\centerline{\bf References}

\item{[\un]}{E. Cremmer, S. Ferrara, L. Girardello and A. Van
Proeyen,
\np{212}{1983}{413}.}
\item{[\deux]}{D. Gross, J. Harvey, E. Martinec and R. Rohm,
\np{256}{1985}{253} and {\bf B267} (1986) 75;\nl
P. Ginsparg, \pl{197}{1987}{139}.}
\item{[\trois]}{V. Kaplunovsky, \np{307}{1988}{145} and ERRATUM
ibid. {\bf B382} (1992) 436;\nl L. Dixon, V. Kaplunovsky and J.
Louis,
\np{355}{1991}{649}.}
\item{[\quatre]}{I. Antoniadis, K.S. Narain and T.R. Taylor,
\pl{267}{1991}{37};\nl
J.-P. Derendinger, S. Ferrara, C. Kounnas and F. Zwirner,
\np{372}{1992}{145} and \pl{271}{1991}{307};\nl
S. Ferrara, C. Kounnas, D. L\"ust and F. Zwirner,
\np{365}{1991}{431};\nl
P. Mayr and S. Stieberger, \np{407}{1993}{725};\nl
V. Kaplunovsky and J. Louis, \np{444}{1995}{191};\nl
D. Bailin, A. Love, W.A. Sabra and S. Thomas, Mod. Phys. Lett.
{\bf A9} (1994) 67 and {\bf A10} (1995) 337.}
\item{[\sept]}{I. Antoniadis, E. Gava and K.S. Narain,
\pl{283}{1992}{209}
and \np{383}{1992}{93}.}
\item{[\huit]}{I. Antoniadis, E. Gava, K.S. Narain and T.R. Taylor,
\np{407}{1993}{706} and {\bf B413} (1994) 162.}
\item{[\cinq]}{E. Kiritsis and C. Kounnas, \np{442}{1995}{472}.}
\item{[\unif]}{I. Antoniadis, J. Ellis, R. Lacaze and D.V.
Nanopoulos,
\pl{268}{1991}{188};\nl
L.E. Ib\'a\~nez, D. L\"ust and G. Ross,
\pl{272}{1991}{251};\nl
L. Dolan and J.T. Liu, \np{387}{1992}{86};\nl
P. Mayr, H.-P. Nilles and S. Stieberger,   \pl{317}{1993}{53};\nl
C. Kounnas, in the proceedings of the {\sl International
Workshop on Supersymmetry and Unification of Fundamental
Interactions,}
Palaiseau, France, 15--19 May 1995;\nl
P. Mayr and S. Stieberger, \pl{355}{1995}{107};\nl
H.-P. Nilles and S. Stieberger, hep-th/9510009;\nl
K.R. Dienes and A.E. Faraggi, \prl{75}{1995}{2646} and
hep-th/9505046;\nl
M. Chemtob, hep-th/9506178.}
\item{[\treize]}{P.M. Petropoulos and J. Rizos, hep-th/9601037.}
\item{[\neuf]}{E. Martinec, \pl{171}{1986}{189}.}
\item{[\dix]}{E. Kiritsis, C. Kounnas and D. L\"ust,
Int. J. Mod. Phys. {\bf A9} (1994) 1361;\nl
C. Kounnas, \pl{321}{1994}{26}.}
\item{[\six]}{E. Kiritsis and C. Kounnas, in the
proceedings of {\sl STRINGS 95,
Future Perspectives in String Theory,} Los Angeles,
 CA, 13--18 March 1995.}
\item{[\onze]}{J. Atick, L. Dixon and A. Sen, \np{292}{1987}{109}.}
\item{[\douze]}{S. Lang, {\sl Elliptic Functions} (Addison-Wesley,
Reading, Mass., 1973); \nl
J.-P. Serre, {\sl Cours d'Arithm\'etique} (PUF, Paris, 1988).}
\item{[\thgen]}{E. Kiritsis, C. Kounnas, P.M. Petropoulos and J.
Rizos,
preprint CERN-TH/96-90, April 1996;\nl
E. Kiritsis, C. Kounnas, P.M. Petropoulos and J. Rizos, in
the proceedings of
the {\sl 5th Hellenic School and Workshops on Elementary Particle
Physics,}
Corfu, Greece, 3--24 September 1995.}
\item{[\quinze]}{L. Dixon, V. Kaplunovsky and J. Louis,
\np{329}{1990}{27}.}
\item{[\seize]}{I. Antoniadis, S. Ferrara, E. Gava, K.S. Narain and
T.R. Taylor, \np{447}{1995}{35}.}
\item{[\dixsept]}{G. Lopes Cardoso, D. L\"ust and T. Mohaupt,
\np{450}{1995}{115};\nl
B. de Wit, V. Kaplunovsky, J. Louis and D. L\"ust,
\np{451}{1995}{53}.}
\item{[\quatorze]}{J.A. Harvey and G. Moore, hep-th/9510182.}
\item{[\sch]}{J.H. Schwarz, hep-th/9512053.}
\item{[\vingt]}{E. Kiritsis, C. Kounnas and P.M. Petropoulos,
work in progress.}

\end